\documentstyle[12pt]{article}

\begin{document}

\def\be{\begin{equation}}
\def\en{\end{equation}}
\def\bear{\begin{eqnarray}}
\def\enar{\end{eqnarray}}
\newcommand{\mn}{\mu\nu}
\newcommand{\rs}{\rho\sigma}

\begin{flushright}
WU-AP/38/94 \\
gr-qc/9406003 \\
June 1, 1994
\end{flushright}

\vskip 1.0cm

\baselineskip .5in

\begin{center}
{\LARGE Finding Principal Null Directions \\
for Numerical Relativists }

\vskip 1.0cm
\baselineskip .35in

{\sc  Laurens Gunnarsen},
{\sc  Hisa-aki Shinkai}\footnote{Electronic address:
shinkai@cfi.waseda.ac.jp}
and {\sc Kei-ichi Maeda}\footnote{Electronic address:
maeda@cfi.waseda.ac.jp}
\vskip 0.5cm

 Department of Physics, Waseda
University,\\ Shinjuku-ku, Tokyo 169,  Japan
\vskip 0.5cm

\end{center}

\vskip 1.0cm

\begin{center}
{\bf abstract}
\end{center}

\baselineskip .3in
We present a new method for finding principal null directions (PNDs).
Because our
method assumes as input the intrinsic metric and extrinsic
curvature of a
spacelike hypersurface, it should be particularly useful to numerical
relativists.
We illustrate our method by finding
the PNDs of the
Kastor-Traschen spacetimes, which contain arbitrarily many
$Q=M$ black holes
 in a de Sitter back-ground. \\


\vfill

\newpage
\baselineskip .30in

\section{Introduction}

According to General Relativity, light propagating from a spherical
source to a distant observer through spacetime curved by the
mass-energy of intervening bodies conveys to that observer a distorted
(i.e., non-circular) image of the source.  The amount of observed
distortion depends on both the direction and the distance to the source,
and of course vanishes as that distance approaches zero. Interestingly,
the distortion growth rate also vanishes in this limit, independently
of the direction from which the observer approaches the source.
We may sum all this up by writing $D={1 \over 2}C(\epsilon, \theta,
\phi) \epsilon^2$, where $D$ is a measure of the image distortion,
$\epsilon$ is a measure of the distance to the source, and the
coefficient
$C(\epsilon, \theta, \phi)$ depends smoothly on $\epsilon$ (even at
$\epsilon=0$).  Remarkably, for certain special directions
($\theta_i, \phi_i$), it happens that $C(0, \theta, \phi)=0$.
In these directions, of which there can be at most four, the
distortion vanishes unusually fast as the observer
approaches the source.
These approximately distortion-free directions are called
 {\it principal
null directions} (PNDs) \cite{PR}.

PNDs provide a great deal of gauge-invariant
information about solutions to Einstein's equation.  Indeed, in vacuum,
they provide nearly as much information as the entire Riemann curvature.
In the early 1960s, many researchers exploited this fact to obtain a
large class of exact vacuum solutions with fewer than four distinct
principal null directions at each point \cite{Kramer}.
  Some of these so-called
algebraically special solutions were easy to interpret.
Those with only one PND at each point, for example,
clearly represented gravitational waves.  But the significance of
algebraically special solutions as a class remained somewhat obscure
until Sachs proved his celebrated `peeling theorem' \cite{peeling}.
This theorem asserts that the PNDs determined by
the gravitational field of a bounded source gradually coalesce as
distance from the source increases.  The greater the distance,  the
more accurately we may approximate the field by ever more `special'
algebraically special solutions.

Sachs's peeling theorem was only the first of many results linking the
behavior of a spacetime's PNDs to the physical
situation that the spacetime models.  More recently, for example,
Arianrhod {\it et. al.} have shown how to infer some
interesting physical
characteristics of the static, cylindrically symmetric Curzon
spacetime from the (admittedly  somewhat complex) behavior of
its PNDs \cite{curzon}.

Despite their proven usefulness in constructing and interpreting exact
solutions to Einstein's equation, PNDs have played
almost no role in work on numerical solutions.
A language barrier seems at least partly responsible for this
unfortunate state of affairs.  Spinors and null tetrads have
traditionally been the tools of choice in work on PNDs, and neither
of these tools fits naturally into the `3+1' framework \cite{adm}
that is the mainstay of numerical relativity.

In this paper, we present a new method for calculating PNDs tailored
to the needs of numerical relativists.  Our method assumes as input
the induced metric and extrinsic curvature of a spacelike hypersurface,
and produces as output the projections into that hypersurface of the
PNDs.

In the next section, we present the details of our method, and
in section 3 we apply it to obtain the PNDs of the Kastor-Traschen
solutions.
The result is a series of pictures analogous to those in Arianrhod {\it
et. al.} showing how the PNDs of various Kastor-Traschen solutions
vary from point to point within the spacelike
hypersurfaces of a natural
slicing.

\section{A `3+1' Method for computing PNDs}
\subsection{Our strategy}

Here, we present our method as it applies to vacuum spacetimes with
cosmological constant  $\Lambda$.
We begin by fixing a triple ($\Sigma, h_{ab}, p_{ab}$), where $\Sigma$
is a smooth 3-manifold, $h_{ab}$ is a Riemannian metric on $\Sigma$,
and $p_{ab}$ is the extrinsic curvature of $\Sigma$.
The constraints on $h_{ab}$ and $p_{ab}$ are
\bear
R-p_{ab}p^{ab}+p^2 &=& 2 \Lambda, \label{hamilt} \\
D_a(p^{ab}-ph^{ab})&=& 0. \label{moment}
\enar
Here $p=p_{ab}h^{ab}$, $D_a$ is the (unique) torsion-free derivative
operator compatible with $h_{ab}$, and $R=R_{ab}h^{ab}$, where
$R_{ab}v^b=-2D_{[a}D_{m]}v^m$ for all smooth $v^a$ on $\Sigma$.

{}From ($h_{ab}, p_{ab}$) we construct two further tensor fields
$E_{ab}, B_{ab}$ as follows:
\bear
E_{ab} &=& R_{ab}-p_a^{~m}p_{bm}+pp_{ab}-{2 \over 3} \Lambda h_{ab}, \\
B_{ab} &=& \varepsilon^{~mn}_aD_mp_{nb},
\enar
where the tensor field $\varepsilon_{abc}=\varepsilon_{[abc]}$
satisfies $\varepsilon_{abc}\varepsilon^{abc}=3!$
It follows from (\ref{hamilt}) and (\ref{moment}) that the fields
$E_{ab}, B_{ab}$ are both trace-free and symmetric.

The next step is to choose a unit vector field $\hat{z}^a$ on
$\Sigma$, and to
decompose $E_{ab}, B_{ab}$ into components along and perpendicular
to $\hat{z}^a$.  We set
\bear
e &=& E_{ab}\hat{z}^a\hat{z}^b, \label{scalarE} \\
e_a &=& E_{bc}\hat{z}^b(\delta_a^{~c}-\hat{z}_a\hat{z}^c),
\label{vectorE} \\
e_{ab} &=& E_{cd}(\delta_a^{~c}-\hat{z}_a\hat{z}^c)
(\delta_b^{~d}-\hat{z}_b\hat{z}^d)+{1
\over 2} e s_{ab}, \label{tensorE} \\
b &=& B_{ab}\hat{z}^a\hat{z}^b, \label{scalarB} \\
b_a &=& B_{bc}\hat{z}^b(\delta_a^{~c}-\hat{z}_a\hat{z}^c),
\label{vectorB} \\
b_{ab} &=& B_{cd}(\delta_a^{~c}-\hat{z}_a\hat{z}^c)
(\delta_b^{~d}-\hat{z}_b\hat{z}^d)+{1
\over 2} e s_{ab}, \label{tensorB}
\enar
where  $s_{ab}=h_{ab}-\hat{z}_a\hat{z}_b$. Finally, we set
\bear
\Psi_0 &=&  (-e_{ab}+ J_a^{~c}b_{bc})m^a m^b, \\
\Psi_1 &=&  {1 \over \sqrt{2}}(e_a-J_a^{~c}b_c)m^a, \\
\Psi_2 &=& {1 \over 2} ( e -i b), \\
\Psi_3 &=&  {1 \over \sqrt{2}}(e_a+J_a^{~c}b_c){\bar m}^a, \\
\Psi_4 &=& (-e_{ab}-J_a^{~c}b_{bc}){\bar m}^a {\bar m}^b.
\enar
where $J_a^{~b} \equiv \varepsilon_a^{~bc}\hat{z}_c $ is
a rotation by 90
degrees in the plane orthogonal to $\hat{z}^a$, and
$m^a  = {1 \over \sqrt{2}}(\hat{x}^a-i \hat{y}^a)$ for some pair of
orthogonal unit vector fields that span that plane.

Given $\Psi_0, \cdots, \Psi_4$, we need only solve the equation
\be
\Psi_4 z^4+4\Psi_3 z^3+6\Psi_2 z^2+4\Psi_1 z +\Psi_0  =0.
\label{quartic1}
\en
to get our final answer: for each root $z_i=\tan {\theta_i \over 2}
 e^{-i\phi_i}~ (i=1,\cdots,4)$,
the unit vector
\be
P_{(i)}^{~~a}=\cos \theta_i \hat{z}^a +
\sin \theta_i \cos \phi_i \hat{x}^a + \sin \theta_i
\sin \phi_i \hat{y}^a \label{pnd}
\en
determines a principal null direction.
By this we mean the following: at points of $\Sigma$ in
the 4-dimensional maximal
evolution ${\cal M}$ of ($\Sigma, h_{ab}, p_{ab}$), $t^a+P_{(i)}^{~~a}$
is a principal null vector for each $i$, where $t^a$  is the unit
normal to $\Sigma$ in ${\cal M}$.

This completes the description of our method. In the following section,
we recall the steps required to solve (\ref{quartic1}).

\subsection{Procedure for finding PNDs}

We follow the d'Inverno-Russel-Clark method \cite{AIJM} to find the
 solutions $z_i$ of (\ref{quartic1}). Upon setting $y=\Psi_4z+\Psi_3$,
(\ref{quartic1}) becomes
\be
y^4+6Hy^2+4Gy+K=0 \label{quartic2},
\en
where $H, G$ and $K$ stand for the following combinations of
the complex numbers $\Psi_0, \cdots,\Psi_4$:
\bear
I &\equiv& \Psi_4\Psi_0-4\Psi_1 \Psi_3 +3 \Psi_2^2, \\
J &\equiv&  \det
 \left| \begin{array}{ccc}
           \Psi_4 & \Psi_3 & \Psi_2 \\
           \Psi_3 & \Psi_2 & \Psi_1 \\
           \Psi_2 & \Psi_1 & \Psi_0 \end{array}
 \right|,\\
K &\equiv& \Psi_4^2 I-3 H^2, \\
H &\equiv& \Psi_4\Psi_2-\Psi_3^2, \\
{\rm and}~~~G &\equiv& \Psi_4^2\Psi_1-3\Psi_4\Psi_3\Psi_2+2\Psi_3^3.
\enar
The solutions of (\ref{quartic2}) are simply expressed in terms of
 the solutions of
\be
\lambda^3-I\lambda+2J=0, \label{quartic3}
\en
which are
 \bear
\lambda_1 &=&   -(          P
 +{              I \over{ 3 P }}), \\
\lambda_2 &=& -(e^{i2\pi/3} P
 +    e^{i4\pi/3}{I\over {3 P }}), \\
\lambda_3 &=& -(e^{i4\pi/3} P
 +    e^{i2\pi/3}{I \over {3 P }})
 \enar
where $P= \{ J+\sqrt{J^2-(I/3)^3}~~\}^{1/3}$.
{}From the $\lambda_i~ (i=1,2,3) $ we now determine three further
 complex numbers,
$\alpha, \beta,\gamma$,  using the following equations:
\bear
\alpha^2&=& 2\Psi_4 \lambda_1-4H,\label{1.8}\\
\beta^2&=& 2\Psi_4 \lambda_2-4H,\\
\gamma^2&=& \alpha^2+\beta^2+4H,\label{1.10}\\
{\rm and}~~~\alpha\beta\gamma&=&4G \label{1.11}
\enar
(\ref{1.8}) - (\ref{1.10}) determine $\alpha, \beta,\gamma$ up to sign,
and (\ref{1.11}) determines the signs.
Then the solutions of (\ref{quartic2}) are given as
${1 \over 2}(\alpha+\beta+\gamma), {1 \over 2}(\alpha-\beta-\gamma),
{1 \over 2}(-\alpha+\beta-\gamma)$ and
${1 \over 2}(-\alpha-\beta+\gamma) $.
Finally, from $\alpha,
\beta,\gamma$, we obtain the following four complex numbers:
\bear
z_1&=&-\{ \Psi_3+{1 \over 2}(\alpha+\beta+\gamma)\}/\Psi_4,
\label{z1} \\
z_2&=&-\{ \Psi_3+{1 \over 2}(\alpha-\beta-\gamma)\}/\Psi_4,\\
z_3&=&-\{ \Psi_3+{1 \over 2}(-\alpha+\beta-\gamma)\}/\Psi_4, \\
z_4&=&-\{ \Psi_3+{1 \over 2}(-\alpha-\beta+\gamma)\}/\Psi_4. \label{z4}
\enar
These numbers are the solutions of (\ref{quartic1}).


\section{PNDs of the Kastor-Traschen solutions}

In this section, we illustrate our method by finding the PNDs of the
Kastor-Traschen (KT) solutions \cite{KT}.
Because these solutions admit no timelike Killing vector field and
contain black holes that undergo rapid relative motion, we would expect
their PNDs to exhibit a variety of interesting and instructive
behaviors.

\subsection{Kastor-Traschen solutions}
The  KT solutions to Einstein's
equation with cosmological constant contain arbitrary many
$Q=M$ black holes that participate in an overall de Sitter expansion
or contraction. In the $\Lambda \rightarrow 0$
limit, the KT solutions reduce to the Majumdar-Papapetrou
solutions, in which the balance between gravitational attraction and
electrostatic repulsion among the black holes causes each to maintain
its position relative to the others eternally.  To write the KT
metric, we first choose $(x_i, y_i, z_i) \in {\bf R}^3$,
$i=1,2,\cdots,N$, and set $ r_i=\sqrt{(x-x_i)^2+(y-y_i)^2+(z-z_i)^2}$.
Then
\be
ds^2=-{1 \over \Omega^2} dt^2+a(t)^2 \Omega^2(dx^2+dy^2+dz^2),
\label{KTmetric}
\en
$$\mbox{where}~~
\Omega=1+\sum_{i=1}^N {M_i \over a r_i},~~ a=e^{Ht}~~~\mbox{and}~~
H=\pm \sqrt{\Lambda \over 3}.$$
Naively, we interpret $M_i$
as the mass of the $i{\rm th}$ black hole, although we have
neither an asymptotically flat region nor event horizons available
to convert this naive interpretation into a rigorous one.

\subsection{PNDs of Kastor-Traschen solutions}

To apply the method of section 2 to the KT solutions, we must first
choose spacelike hypersurfaces $\Sigma$ and orthonormal triads
$\hat{x}^a, \hat{y}^a, \hat{z}^a$
 tangent to these $\Sigma$.  The natural choice would
seem to be the  $t= const.$ surfaces for $\Sigma$, and
\be
\hat{x}^a={1 \over a \Omega}
\left( {\partial \over \partial x}\right) ^a,~~
\hat{y}^a={1 \over a \Omega}
\left( {\partial \over \partial y}\right) ^a,~~
\hat{z}^a={1 \over a \Omega}
\left( {\partial \over \partial z}\right) ^a
\en
for the triad.  We keep this choice in force throughout
in what follows.

We now select various values of the free parameters ($x_i, y_i, z_i$)
and $M_i$ appearing in the KT metric (\ref{KTmetric}), and
plot the PNDs our method generates.  For simplicity, we take all
$z_i=0$ --- that is, we confine all black holes to the $x-y$
coordinate plane --- and plot the projections of the PNDs into that
plane.

For the case of one black hole, the KT solution is just
the Reissner-Nordstr\"om-de
Sitter solution in cosmological coordinates, and its horizons are
located at $r_\pm=(1-2M|H|\pm\sqrt{1-4M|H|})/2a|H|$,
where $r_-$ and $r_+$ denote the
 outer black hole horizon and  de Sitter horizon, respectively.

Figure 2  shows the PNDs of a KT solution containing two black holes
of equal mass ($M_1=M_2=1.0$) separated
by a coordinate distance $d=5$.  In this case,
$r_+=7.87M$ and $r_-=0.127M$, so that
the size of each black hole
is the same as black points indicating its location.
Four PNDs at each point are presented using arrows,
but because these four PNDs
coincide in pairs, with the coincident pairs pointing in opposite
directions, we get a picture that looks like the electric field lines
of equal charges. This indicates that the spacetime is Petrov type D
at all points.

As we decrease the separation between two black holes, we expect that
at a certain critical value of that separation, their horizons
will coalesce.  When two equal-mass black holes inhabit a contracting
 ($H<0$) KT spacetime (total mass $M$), the numerical results of
\cite{hoop-conj} suggest a value of $d\sim 0.45M$
 for this critical separation,
which leads us to believe that the two black holes in Figure 2
have distinct horizons.

In Figure 3, we show  the case of two black holes with different
masses ($M_1:M_2=1:10$) separated by (a) $d=5$, (b) $d=2$
and (c) $d=0.5$ in a contracting background ($H<0$).
Since the black holes move along the natural
trajectories in that background, we may interpret these figures as
snapshots of coalescing black holes.
These cases have none of the symmetry of  Figure 2,
and as a result, the field of PNDs is algebraically general (Petrov
type I).
However, we also see that the spacetime near each black hole is
almost Petrov type D, as is a single black hole,  and that
at great distances from the holes the PNDs exhibit the same almost
type-D behavior.  This is just as we would expect.
Moreover, we see that the PNDs on the axis connecting
the two black holes exhibit Petrov type D behavior.
Although we have no value for the critical separation in the case
of two black holes with different masses, Figure 3(c) suggests that
as the separation tends to zero, the spacetime's Petrov type approaches
that of a single black hole.


We can also draw the pattern of PNDs for KT solutions containing
many black holes. As an example, we show in Figure 4 the PNDs of
three black holes
of equal mass located  at the vertices of an equilateral
triangle.

\section{Conclusions and discussion}

The results of the previous section illustrate the advantages of
plotting PNDs over other methods of assessing how nearly algebraically
special a spacetime is.  Particularly when the spacetime is known only
approximately, other methods are hard to apply, for they require us to
determine whether certain scalar combinations of curvature components
(e.g., $I^3-6J^2$) are exactly zero or not.

We hope the method presented in this paper will help numerical
relativists to deal with at least two subtle issues.
First, there is the problem of how to set
initial data representing bounded sources free of incoming radiation.
Because Sachs's peeling theorem assumes no incoming radiation,
any failure of the PNDs computed from ($\Sigma, h_{ab}, p_{ab}$)
to exhibit the proper peeling
behavior shows that ($\Sigma, h_{ab}, p_{ab}$) is an unsatisfactory
model of the
instantaneous state of an isolated system.  Our method thus
allows us to recognize and reject initial data cluttered with
spurious incoming radiation.  This should be of some help.
Second, there is the problem of detecting spurious gravitational waves
in numerically generated spacetimes  that result from physically
inappropriate boundary conditions.
We  hope that such waves will manifest themselves in a pathological
time dependence of the PNDs near numerical boundaries.

\vskip 0.3cm
{\bf Acknowledgment}

This work was supported partially by the Grant-in-Aid
for Scientific Research Fund of the Ministry of Education, Science and
Culture Nos. 06302021 and  06640412.
\par

\newpage

\newpage
\baselineskip .3in
\vskip 0.5cm
\begin{center}
{\bf Figure Captions}
\end{center}
\vskip 0.5cm

\noindent
Figure 1:  \\
\noindent
The solutions $z_k$  (\ref{z1})-(\ref{z4})
for each $k=1,...,4$ at each point determine the PNDs.
If we write $z_k$ as $z_k=r_k e^{-i\phi_k}$, and set
$r_k=\tan {\theta_k \over 2}$, then each PND  $P_{(k)}^{~~a}$
points in the direction
$(\theta_k, \phi_k)$ indicated by the bold arrow in the figure,
which we write as in (\ref{pnd}).

\vskip 0.5cm
\noindent
Figure 2:  \\
\noindent
PNDs of the Kastor-Traschen solution for the case of two black holes
of equal mass ($M_1=M_2=1$) separated by a coordinate distance $d=5$.
Black holes are located at ($x,y$)=(3.0, 5.5) and (8.0, 5.5), and the
arrows indicate PNDs at each point.

\vskip 0.5cm
\noindent
Figure 3:  \\
\noindent
PNDs of a KT solution containing two black holes with different
masses ($M_1:M_2=1:10$) separated by (a) $d=5$, (b) $d=2$
and (c) $d=0.5$.
In these cases,  the symmetry of Figure 2 is absent, we see
that the field is algebraically general (Petrov type I).
We also see, however, that the spacetime near each black hole  and
along the axis joining the two black holes is almost  Petrov type D,
 and that in regions far from both holes, the spacetime is
also almost Petrov type D, like a single black hole.

\vskip 0.5cm
\noindent
Figure 4:  \\
\noindent
PNDs of a KT solution with three black holes
of equal mass  at the vertices of an equilateral triangle.
Black holes are located at
($x,y$)=(2.5, 3.0), (8.5, 3.0) and (5.5, 8.2).

\end{document}